\documentclass[12pt]{article}
\usepackage{amsmath, amsthm, amssymb,bbm}

\title{\bf Noise induced current in a double-well trap}

\author{Sebastiano Anderloni$^{a,b}$, Fabio Benatti$^{a,b}$, 
Roberto Floreanini$^{b}$\\
\small ${}^a$Dipartimento di Fisica Teorica, Universit\`a di Trieste,
Strada Costiera 11, 34014 Trieste, Italy\\
\small ${}^b$Istituto Nazionale di Fisica Nucleare, Sezione di Trieste,
34100 Trieste, Italy}

\date{\null}

\begin{document}

\maketitle

\begin{abstract}
\noindent
We study the behavior of cold atoms trapped in optical double well potentials
in presence of noise, either generated by an external
environment or by the trap itself. We show that quite in general 
the noise can induce a current between the two wells even when tunneling
between the two wells is highly suppressed (Mott insulator phase).
An engineered environment could provide a test ground for the behavior of such a current.
\end{abstract}

\vskip 1 cm

\section{Introduction}

Cold atoms (typically Bose-Einstein condensates) in optical lattices provide a 
very manageable and versatile tools for experimental tests 
in many-body quantum physics and as such have recently attracted a lot of interest.%
\footnote{The literature on this topic is enormous; for a recent review see \cite{lewe}
and the long list of references therein.}
In a suitable approximation, the dynamics of bosonic atoms in this kind of traps can be described by a Bose-Hubbard model \cite{cirac}, which, depending on the value of the ratio between the hopping amplitude and the on site repulsive boson-boson interaction, predicts a superfluid and a Mott insulator phase. Much of the
experimental work on such systems have been dedicated to the study of this phase transition.

These systems are usually treated as isolated. However this is
only an approximation, and there are instances when the presence of a weakly coupled external environment 
can not be neglected. Typically, this happens when the
optical lattice is immersed in a thermal bath, either external or formed by the fraction
of cold atoms that are not in the condensed phase, or when the lattice itself is subjected to stochastic noise.
In these cases, the condensate gas in the lattice need to be treated as an open quantum system, and its
dynamics is altered by effects of noise and dissipation \cite{al,gorini1,spohn,alicki,petruccione,bf}.

In the following, we shall study such effects in the particular case of a lattice with just two sites, 
where the cloud of condensed bosonic atoms is confined in a double-well potential (extension to the
multiple sites case will be reported elsewhere). In the ideal, isolated situation,
the Mott insulator phase corresponds to a blocked regime where the occupation numbers of the 
two wells is fixed and no atom is allowed to cross the barrier passing from a well to the other. 
In particular, no current across the trap barrier is observed.

On the other hand, when the trap can not be considered isolated, the two wells can be effectively connected
through the indirect action of the environment, whose presence may then break the insulating regime.
This is in fact what it is found by analyzing in detail the time evolution of the current operator:
as soon as the interaction with the environment is switched on, an environment induced current is
generated via a purely noisy mechanism. Through an engineered environment, 
{\it e.g.} by injecting and modulating a stochastic noise in the double well potential, this
current can be in principle experimentally analyzed.

%\vskip 1cm 

\section{Double-well trap}

As explained above, we shall study the behavior of a system of condensed bosonic atoms trapped 
in a double well potential, weakly coupled to an external environment. 
In a suitable approximation, the dynamics of the system
alone ({\it i.e.} in absence of the environment) is very well described by a Bose-Hubbard type hamiltonian
\cite{cirac}, 
that in the present case takes the simple form
\begin{equation}
\label{BH}
H_{S}=-T(a^\dag_{1}a_{2}+a^\dag_{2}a_{1})+U(n_{1}^{2}+n_{2}^{2})
+\varepsilon_{1} n_{1}+\varepsilon_{2} n_{2}\ ,
\end{equation}
where the operators $a_{i}$, $a^\dag_{i}$ , $i=1,2$ are annihilation and creation operators 
for atoms in the two wells, obeying the standard oscillator algebra 
$[a_{i},\ a^\dag_{j}]=\delta_{ij}$, with $n_{i}=a^\dag_{i}a_{i}$, the number operators.
The various contributions in the hamiltonian (\ref{BH}) are characterized by their physical meaning: 
the first is an hopping term dependent on the tunneling
amplitude $T$, while the remaining are on-site energy terms; 
the one proportional to the coupling constant $U$ is
quadratic in the number operator and is due to the boson-boson repulsive contact interaction, 
while the second one is due to the trapping potential and may be different for each well.
The parameter $\varepsilon_{i}$ represents the energy of the bottom 
of the $i$-th well and therefore describes its depth. 

As well known \cite{lewe}, the hamiltonian (\ref{BH}) describes the quantum phase transition 
between a superfluid and insulator phases, whose order parameter is given by the ratio $T/U$. 
We are interested in studying the regime where the atomic cloud is well separated in the two wells; 
this is the case when the hopping energy is very small compared to the on-site one. 
Neglecting the shifts $\varepsilon_{i}$ which are of the same order for both the wells, 
when an atom hops, it loses an energy $T$ and gains instead an energy $U$; when $T/U\ll 1$, 
this situation is energetically suppressed and, at equilibrium, the ground state of the system is given by
a Fock state with equal number $N$ of bosons per well, $|N,\, N\rangle\equiv |N\rangle\otimes|N\rangle$. 
This is the typical situation
of a Mott insulator, where no current is flowing between the two wells.

In order to give a precise meaning to this statement, one needs to introduce a suitable
current operator. Following \cite{kolo,buch}, we shall first consider the barycenter of the system of atoms
in the two wells; in suitable units, it is simply given by
\begin{equation}
Z=\frac{(a^\dag_{1}a_{1}-a^\dag_{2}a_{2})}{2N}\ .
\end{equation}
The velocity with which the barycenter moves can be obtained taking the time derivative of this operator:
\begin{equation}
	\label{}
	{dZ\over dt}=i[H_{S},\ Z]=\frac{iT}{2N}(a^\dag_{1}a_{2}-a^\dag_{2}a_{1})\ .
\end{equation}
The current operator describes the flow of atoms between the two wells and thus
it should be proportional 
to this velocity operator; in the chosen units, it can be simply taken to be
\begin{equation}
	\label{J}
	J\equiv i(a^\dag_{1}a_{2}-a^\dag_{2}a_{1})\ .
\end{equation}

In the ground state $|N,\, N\rangle$, this operator have vanishing mean value, meaning that no current flows across the barrier when the wells are equally filled. Moreover it is easy to check that also its time derivative vanishes:
	\begin{equation}
	\label{dotJ}
	\frac{d}{dt}\langle J\rangle_N=\Big\langle i[H_{BH},\ J]\Big\rangle_N=0\ .	
	\end{equation}
As we shall see, this conclusion is in general no longer true in presence of an environment.

%\vskip 1cm

\section{Noise induced current}

When the interaction with the environment $E$ is switched on, the hamiltonian describing
the evolution of the total system, condensed gas in the trap plus environment, can be decomposed as
	\begin{equation}
	\label{HSE}
	H_{SE}=H_{S}+H_{E}+\lambda H_{I}\ ,
	\end{equation} 
where $H_{S}$, driving the free motion of the atoms in the double well trap, is as in (\ref{BH}), 
$H_{E}$ describes the evolution of the  environment alone, 
while $H_{I}$ takes care of the interaction between atoms and environment, 
$\lambda$ being a small coupling constant, $\lambda\ll1$. For a weakly coupled environment,
the form of $H_I$ can be taken to be bilinear in the system and environment variables,
\begin{equation}
\label{HI}
H_{I}=\sum_{i=1}^{4}V_{i}\otimes B_{i}\ ,
\end{equation}
where $B_i$ are hermitian operators describing environment observables, while the system operators $V_i$
are given by the hermitian combinations of the oscillator variables:
\begin{equation}
\label{V}
V_i=\{(a_{1}+a^\dag_{1}),\ i(a_{1}-a^\dag_{1}),\ (a_{2}+a^\dag_{2}),\ i(a_{2}-a^\dag_{2})\}\ .
\end{equation}

The state of the composite system $S+E$, 
described in general by a density matrix $\rho_{SE}$, evolves unitarily with the hamiltonian (\ref{HSE}). However, the dynamics of the density matrix $\rho\equiv {\rm Tr}_E[\rho_{SE}]$ describing the state of the
condensed atoms in the trap and obtained by averaging over
the environment degrees of freedom, is in general complicated, containing non-linearities, secular
terms and memory effects \cite{al,gorini1,spohn,alicki,petruccione,bf}. When system and environment are initially uncorrelated, no secular terms
or non-linearities arise; this is a rather common situations in actual experimental realizations,
so that we shall henceforth assume that at the initial time $t=\,0$, 
the state of the composite system be in factorized form, $\rho_{SE}(0)=\rho(0)\otimes\rho_E$, 
$\rho_E$ being the reference equilibrium state of the environment, {\it e.g.}
a thermal state $\rho_{E}\simeq e^{-\beta H_{E}}$.

Memory effects are typically short time phenomena, due to the initial interaction
between subsystem and environment. We are not interested in the details of such phenomena;
rather, we would like to study the effective dynamics of the subsystem alone
on long time scales, for which memory effects have already died out. 
Operatively, since the noise and dissipation due to the environment
start to become relevant on time scales of order $\lambda^{-2}$, this is obtained by passing
to a suitable rescaled time variable, $t\to t/\lambda^{2}$ and then letting
$\lambda\to 0$. This procedure is known as ``Markovian approximation'' and
allows disregarding the initial memory effects. In order to be physically applicable,
this approximation requires an {\it a priori} unambiguous separation between
subsystem $S$ and environment $E$. In the weak coupling regime, this is typically
achieved when the
ratio $\tau_E/\tau_R$, between the typical decay time of correlations in the environment and
the characteristic relaxation time of the subsystem immersed in the environment, is small.

Two different Markovian limits have been discussed in the literature, depending on the way of rendering
the ratio $\tau_E/\tau_R$ small. In the {\it weak coupling limit} \cite{davies1,davies2}, the correlations in
the environment
\begin{equation}
	\label{G}
G_{ij}(t)=\langle B_{i}(t)B_{j}\rangle\equiv{\rm Tr}\Big[B_{i}(t)\, B_{j}\, \rho_E\Big]\ ,\quad
B_{i}(t)=e^{itH_E} B_i\, e^{-itH_E}\ ,
	\end{equation}
are non-singular and assumed to decay fast enough,
\begin{equation}
	\label{}
	\int_{0}^{\infty}{\rm d} t|G_{ij}(t)|(1+t)^{\epsilon}<\infty\ ,
	\end{equation}
with $\epsilon$ a positive constant. For typical environments, like heat baths, one finds
an exponential behavior:
\begin{equation}
\label{Gexp}
G_{ij}(t)\simeq G_{ij}\ e^{-\mu_E\, t}\ , \qquad \mu_E=1/\tau_E\ ,
\end{equation}
with $G_{ij}$ a constant hermitian matrix.
In the {\it singular coupling limit} \cite{gorini2} instead, it is the decay time of correlations in
the environment that becomes small, $\tau_E\to 0$; concretely, this is obtained by letting 
the correlation functions tend to a Dirac-delta:
\begin{equation}
G_{ij}(t)\to G_{ij}\, \delta(t)\ .
\label{Gdelta}
\end{equation}
This situation is encountered in a bath with infinitely large temperature or in the case
in which the subsystem is subjected to stochastic noise; in the latter case the positive constants
$G_{ij}$ form a symmetric matrix.

In both cases, the master equation describing the evolution of the subsystem density matrix $\rho(t)$
takes the form
\begin{equation}
\label{ME}
{\partial\rho(t)\over\partial t}=\mathbb L[\rho(t)]
\equiv-i[H_{S}+H^{(2)},\ \rho(t)]+\mathbb D[\rho(t)]\ .
\end{equation}
The environment modifies the standard Liouville-von Neuman equation for  $\rho$ 
by an effective correction $H^{(2)}$ to the free hamiltonian $H_S$ 
and a pure dissipative contribution $\mathbb D$ which cannot be written in hamiltonian form.
Explicitly, these new contributions take the standard Kossakowski-Lindblad expression 
\cite{kossa1,kossa2,lindblad}:
	\begin{eqnarray}
	\label{D}
	&&\mathbb D[\rho]=\lambda^2\, \sum_{\omega
		}\sum_{ij=1}^{4}c_{ij}(\omega)\left[V_{j}^{\dag}(\omega)\rho V_{i}(\omega)
		-\frac{1}{2}\left\{V_{i}(\omega)V_{j}^{\dag}(\omega),\rho\right\}\right]\ , \\
	&&H^{(2)}=\lambda^2\, \sum_{\omega
		}\sum_{ij=1}^{4}s_{ij}(\omega)V_{i}(\omega)V_{j}^{\dag}(\omega)\ , 
	\label{H2}
\end{eqnarray} 
where the formal sum over $\omega$ represents the sum over all possible energy 
differences within the spectrum of the subsystem hamiltonian $H_{S}$.
The two hermitian matrices, $c_{ij}(\omega)$ and $s_{ij}(\omega)$, 
are the Fourier and Hilbert transform of the environment correlation functions, 
respectively, ($\mathcal P$ denotes principal value),
	\begin{eqnarray}
	\label{c}
	&&c_{ij}(\omega)=\int_{-\infty}^{+\infty}{\rm d} t\,  e^{-i\omega t}\, G_{ij}(t)\ , \\
	&&s_{ij}(\omega)=\frac{1}{2\pi}
	\mathcal P\int_{-\infty}^{+\infty}{\rm d} w\, \frac{c_{ij}(w)}{w-\omega}\ .
\label{s}
\end{eqnarray}
They embody the effects of noise and dissipation induced by the environment.
Notice that the Kossakowski matrix $c_{ij}(\omega)$ turns out to be positive, a key requirement
in order to guarantee the positivity of $\rho(t)$ for any time, and thus the physical
consistency of the evolution generated by (\ref{ME}) \cite{al,bf}.
Finally, the Kraus operators $V_{i}(\omega)$ are suitable sandwiches of the system operators 
$V_{i}$ of (\ref{V}) between eigenprojectors of the system hamiltonian $H_S$
(see the following for explicit expressions).

By exponentiation, the operator $\mathbb L[\cdot]$ generates  a family of completely positive and trace-preserving maps $t\to\gamma_{t}=e^{t\mathbb L}$ which constitutes a so-called \textit{quantum dynamical semigroup}  \cite{al,gorini1}, for which: $\gamma_{t}\circ\gamma_{s}=\gamma_{t+s},\ t,s\ge0$,
thus encoding the intrinsic irreversibility of the dynamics.

As discussed before, we are interested in analyzing the behavior of the current operator (\ref{J})
in presence of the environment. Its evolution in time is generated by a master equation
which is obtained from (\ref{ME}) by duality:
\begin{equation}
\label{DME}
{\partial J(t)\over \partial t}=\mathbb L^*[J(t)]\equiv
i[H_{S}+H^{(2)},\ J(t)]+\mathbb D^{*}[J(t)]\ ,
\end{equation}
where the operator $\mathbb L^{*}[J]$ is defined via the duality relation 
Tr$(\mathbb L[\rho]\, J)=\text{Tr}(\rho\, \mathbb L^{*}[J])$, and similarly for
$\mathbb D^*$. 
We shall now study in detail the behaviour of the current operator $J(t)$ in presence of noise
as described by the two regimes of weak and singular coupling limit.

%\vskip .5cm
\subsection{Weak coupling limit}

As discussed before, we are considering the situation in which, prior to the coupling with
the environment, the double well trap is in the Mott insulator
phase, in which tunnelling as induced by the first term in the system hamiltonian
(\ref{BH}) is suppressed, {\it i.e.} $T\ll U$. 
The noise contributions generated by
the presence of the environment are second order in the interaction hamiltonian $H_I$
and therefore turn out to be of order $\lambda^2$ ({\it cf.} the expressions in (\ref{D}), (\ref{H2})); in evaluating the coefficients (\ref{c}), (\ref{s}) 
and the energy differences $\omega$ in the sums (\ref{D}), (\ref{H2}),
one can then safely neglect tunneling contributions, being
subdominant, of order $\lambda^2\, T$. In this approximation, the spectrum of the system
hamiltonian (\ref{BH}) results quadratic in the
the occupation numbers of the two wells, while the corresponding energy differences
can be labelled by an integer $n$, 
\begin{equation}
\omega\equiv\omega_n=\varepsilon +U+2U\, n\ .
\end{equation}
For simplicity, we are focusing on a trap formed by wells with equal depth,
$\varepsilon_1=\varepsilon_2=\varepsilon$;
we will comment on the case $\varepsilon_1\neq \varepsilon_2$ later on.

Explicit evaluation allows casting the dissipative contributions to the dual master equation
(\ref{DME}) in the form:
\begin{eqnarray}
	\label{DW}
&&\mathbb D^*[J]=\lambda^2 \sum_{n}\sum_{ij}h_{ij}(\omega_{n})\left[{\mathcal V}_{i}(\omega_{n})\, J\, 
{\mathcal V}_{j}^\dagger(\omega_{n})-\frac{1}{2}\left\{{\mathcal V}_{i}(\omega_{n}){\mathcal V}_{j}^{\dag}(\omega_{n}),J\right\}\right]\ , \\
&&H^{(2)}=\lambda^2 \sum_{n}\sum_{ij} k_{ij}(\omega_{n}){\mathcal V}_{i}(\omega_{n}){\mathcal V}_{j}^{\dag}(\omega_{n})\ ;
\label{H2W}	
	\end{eqnarray}
the Kraus operators are now given by
	\begin{equation}
	\label{}
	{\mathcal V}(\omega_{n})=\left(P_{n}a_{1}\otimes{\mathbbm 1},\ a^\dag_{1} P_{n}\otimes{\mathbbm 1},\ 
	{\mathbbm 1}\otimes P_{n}a_{2},\ {\mathbbm 1}\otimes a^\dag_{2}P_{n}\right)\ ,
\end{equation}
where $P_n$ are projector operators on single well Fock states with occupation number $n$,
while the Kossakowski matrix takes the form
\begin{equation}
h(\omega_{n})=
	\begin{pmatrix}
	h_{11}(\omega_{n}) & 0 & h_{13}(\omega_{n}) & 0 \\
	0 & h_{22}(-\omega_{n}) & 0 & h_{24}(-\omega_{n}) \\
	h_{13}^{*}(\omega_{n}) & 0 & h_{33}(\omega_{n}) & 0 \\
	0 & h_{24}^{*}(-\omega_{n}) & 0 & h_{44}(-\omega_{n})
\label{h}
\end{pmatrix}\ .
	\end{equation}
The entries of this matrix are linear combinations of the Fourier 
transforms of the correlation functions given in (\ref{c}); 
omitting for simplicity the $\omega_n$ dependence, one has:
\begin{eqnarray}
\nonumber
&&h_{11}=c_{11}+c_{22}+i(c_{21}-c_{12})\\
\nonumber
&&h_{13}=c_{13}+c_{24}+i(c_{23}-c_{14})\\
&&h_{22}=c_{11}+c_{22}+i(c_{12}-c_{21})\\
\nonumber
&&h_{24}=c_{13}+c_{24}+i(c_{14}-c_{23})\\
\nonumber
&&h_{33}=c_{33}+c_{44}+i(c_{43}-c_{34})\\
\nonumber
&&h_{44}=c_{33}+c_{44}+i(c_{34}-c_{43})\ .
\label{hc}
\end{eqnarray}
The matrix $k_{ij}$ has a similar structure in terms of the Hilbert transform of the environment
correlation functions; it will not be relevant for the considerations that follow and thus,
for simplicity, its explicit expression is omitted.

Assume now the system be prepared in the Mott insulator phase, with initial state 
$\rho_N=|N,N\rangle\langle N, N|$;
as already observed, the average of the current operator (\ref{J}) in this state vanish:
$\langle J(0)\rangle_N=\,0$. In order to see whether the presence of the environment alters
this situation, it is sufficient to study the time derivative of the average $\langle J(t)\rangle_N$
at $t=\,0$ as given by the evolution equation (\ref{DME}) with (\ref{DW}) and (\ref{H2W}).
The explicit computation gives:
\begin{equation}
{\partial \langle J(t)\rangle_N\over \partial t}\bigg|_{t=\,0}=\langle\mathbb D^{*}[J(0)]\rangle_N\ ,
\label{Jdot}
\end{equation}
since all hamiltonian contributions vanish, with
\begin{eqnarray}
\nonumber
	\langle\mathbb D^{*}[\hat J(0)]\rangle_N&=&i\lambda^2\Big[(N+1)^{2}\big(h_{13}(\omega_{N})
	-h_{31}(\omega_{N})\big)+N^{2}\big(h_{42}(-\omega_{N-1})-h_{24}(-\omega_{N-1})\big)\Big] \\
	&=& 2\lambda^2\Big[(N+1)^{2}\Im m\big(h_{31}(\omega_{N})\big)
	+N^{2}\Im m\big(h_{24}(-\omega_{N-1})\big)\Big]\ .
\end{eqnarray} 
Therefore, as soon as the interaction with the environment is switched on, a current starts
flowing between the two wells; it has a purely noisy origin and its magnitude is determined
by the elements of the Kossakowski matrix (\ref{h}), 
{\it i.e.} by the correlation functions in the environment.
In the case of an heat bath with correlation functions of exponential type as in (\ref{Gexp}), 
one finds that, for small times and large enough $N$, the current behaves as:
\begin{eqnarray}
	\label{}
\langle J(t)\rangle_N\simeq{\partial \langle J(0)\rangle_N\over \partial t} \ t &=&
{8\, \lambda^2 N^2 \mu_E \over \mu_E^2 + \omega_N^2}\ \Re e\big(G_{14}-G_{23}\big)\ t\\
&\simeq& {2\lambda^2\mu_E\over U^2}\ \Re e\big(G_{14}-G_{23}\big)\ t\ .
\label{JW}
\end{eqnarray}
The intensity of the current clearly depends on the strength $G_{ij}$ 
of the environment correlation functions. In particular, 
the current vanishes if the coupling of the two wells to the environment
is realized through the same operators, {\it i.e.} if the environment
acts exactly in the same way on the two wells; indeed, in this case the r.h.s. of (\ref{JW}) is zero
since $B_{1}=B_{3}$ and $B_{2}=B_{4}$, and therefore, being $G_{ij}$ hermitian,
$\Re e(G_{14})=\Re e(G_{23})$.

Finally, let us briefly consider the situation in which the two wells have initially different depth,
{\it i.e.} $\varepsilon_{1}\neq\varepsilon_{2}$ in the starting hamiltonian (\ref{BH}). The weak coupling limit
procedure involves an average over the microscopic fast motion of the system as generated by (\ref{BH});
this is perfectly justified, since we are focusing on the dynamics over rescaled, long time intervals (this
is usually referred to as ``rotating wave approximation'' \cite{al,gorini1}). When $\varepsilon_{1}\neq\varepsilon_{2}$,
this average procedure gives rise to a simplified Kossakowski matrix, where only the diagonal
elements are non vanishing. In other terms, everything goes as if the two wells interact 
independently with the environment and no correlations between them could be created through
noisy effects, or, more precisely, these correlations average to zero if they are observed
on time intervals rescaled by $\lambda^2$. As a consequence, since the environment can not connect
the two wells, the system stays in the insulator phase also after coupling to the environment and 
the average value of the current operator remains zero.

%\vskip .5cm

\subsection{Singular coupling limit} 

As mentioned before, this Markovian approximation is appropriate
when our double well trap is immersed in an environment
whose correlation functions decay very fast, as happens for instance
in a heat bath at infinitely large temperature, or when the trap is subjected to stochastic noise. 
In practice, one assumes the fluctuations
in the bath to be delta-correlated as given in (\ref{Gdelta}), so that the dissipative contributions
to the master equation (\ref{ME}) become independent from the system hamiltonian (\ref{BH}) \cite{gorini2}. 
The sum over the spectral parameter $\omega$ disappears from the expressions of $\mathbb D$ and $H^{(2)}$
in (\ref{D}) and (\ref{H2}), 
while the Kossakowski matrix becomes constant, $c_{ij}=G_{ij}$, and Kraus operators
$V_i$ as in (\ref{V}). Notice that in general all the entries of $c_{ij}$ are now non vanishing,
so that the structure of the dissipative contributions in (\ref{ME}) is in general richer than in
the weak coupling limit.%
\footnote{ This is
a consequence of the fact that no ``rotating wave approximation'' is necessary
in the singular coupling limit.}

The dual equation, generating the dissipative dynamics for the current operator, takes again the
form (\ref{DME}), where the dissipative contributions can be cast in the form similar to (\ref{DW})
and (\ref{H2W}); explicitly:
	\begin{eqnarray}
	\label{}
	&&\mathbb D^*[J]=\lambda^2 \sum_{ij}h_{ij}\left[{\mathcal V}_{i}\, J\, {\mathcal V}_{j}^{\dag}
	-\frac{1}{2}\left\{{\mathcal V}_{i}{\mathcal V}_{j}^{\dag},\, J\right\}\right]\ , \\
	&&H^{(2)}=\lambda^2 \sum_{ij} k_{ij} {\mathcal V}_{i}{\mathcal V}_{j}^{\dag}\ ,
	\end{eqnarray}
where now one simply have
	\begin{equation*}
	\label{}
	{\mathcal V}=\left(a_{1},\ a^\dag_{1},\ a_{2},\ a^\dag_{2}\right)\ .
	\end{equation*}

The entries of the new Kossakowski matrix $h_{ij}$ are linear combinations of the old one $c_{ij}$,
with relation analogous to those in (\ref{hc}); similar results hold also for the hamiltonian
contributions $s_{ij}$.

In order to see whether also in this case a current can be generated by purely dissipative
effects, one first prepares the double well trap in the Mott insulator phase, described
by the equilibrium state $\rho_N=|N,N\rangle\langle N,N|$; then, one switches on the interaction
with the environment and looks again at the behavior of the average $\langle J(t)\rangle_N$
for small times. Here again only the dissipative part $\mathbb D^{*}$ contributes so that 
the result (\ref{Jdot})
is still valid, while the explicit evaluation gives:
\begin{equation}
\langle J(t)\rangle_N\simeq{\partial \langle J(0)\rangle_N\over \partial t} \ t =
2\lambda^2\Big[(2N+1)\, \Im m(G_{31}+G_{42})+\Re e(G_{14}-G_{23})\Big]\ t\ .
\label{JS}
\end{equation}
Also in the case of delta-correlated environments, a current is initially generated by a purely noisy
mechanism; its intensity depends again on the entries of the Kossakowski matrix,
{\it i.e.} on the strength of the correlations in the bath.

From the expression in (\ref{JS}), one sees that as in the case of the weak coupling limit,
the current vanishes when the environment couples in the same way to the two wells
(as before $\Re e(G_{14})=\Re e(G_{23})$, while $G_{14}\equiv G_{11}$, $G_{42}\equiv G_{22}$
are real, again by the hermiticity of $G_{ij}$).
On the other hand, the mechanism generating the noisy current contribution
is independent from the details of the microscopic system dynamics given
by the hamiltonian (\ref{BH}); this implies that a delta-correlated noise can in principle
generate a current even for $\varepsilon_{1}\neq\varepsilon_{2}$,
{\it i.e.} when the two wells are initially at different depths.

%\vskip 1cm

\section{Outlook}

An optical double-well trap filled with
condensed bosonic atoms immersed in an external, weakly coupled environment
represents an example of an open quantum system. The presence of the environment
generates noise and dissipation, and as a consequence, 
the evolution in time of the atoms in the trap is no longer reversible.
Rather, it is described by a quantum dynamical semigroup, 
generated by a master equation of the Kossakowski-Lindblad
form (\ref{ME}), with (\ref{D}), (\ref{H2}).

As discussed above, the detailed structure of such evolution
depends crucially on the characteristic properties of the environment,
specifically on the behaviour of its two-point correlation functions.
Nevertheless, as a general result, the dissipative dynamics allows
for an indirect coupling between the atoms in separate wells, through
the noisy action of the environment.
Therefore, even if the trap is initially prepared in the Mott insulator phase,
where tunneling is not allowed, the subsequent
coupling to the environment may allow a transitions between the two wells.

This possibility has been studied by analyzing the behaviour 
of a suitably defined current operator, signaling an unbalanced
flow between the two wells. As soon as the interaction with
the environment is switched on, the current starts developing
a non-vanishing, purely dissipative contribution to its average.
This contribution can be isolated by preparing the trap in the
insulator phase, where the current has a vanishing expectation value.
The interaction with the environment will then have the net result
of effectively ``breaking'' the insulator regime.

This effect may be experimentally ascertained through the use of
an engineered environment. The idea is to inject stochastic noise
into the optical trap, operation that, in the formalism developed above,
can be described by adding to the
system hamiltonian (\ref{BH}) an external Gaussian stochastic potential.
One can show \cite{gorini2, bf} that the average over the noise produces
effects similar to those obtained through the coupling with
a delta-correlated environment, and thus, as in (\ref{JS}) a non vanishing dissipative
contribution to the current operator.
By releasing the trap potential right after the source of stochastic noise
has been switched on, one should then be able to measure an unbalance 
in the barycenter position of the ballistic expanding cloud of condensed atoms,
a clear signal of flow of atoms between the two wells.

%\vfill\eject

\end{document}